\documentclass[prf,superscriptaddress,showpacs,longbibliography,aps]{revtex4-1}

\usepackage[final]{graphicx}
\usepackage{amsmath}
\usepackage{verbatim}
\usepackage{color}
\usepackage{lineno}
\usepackage{amssymb}
\usepackage{ulem}
\usepackage{natbib}
\usepackage{textcomp}

\begin{document}

\title{Spatial organization of biomass controls intrinsic permeability of porous systems}

\author{Wenqiao Jiao}
\affiliation{Institute of Earth Science, University of Lausanne, Lausanne 1015, Switzerland}
\affiliation{Dipartimento di Ingegneria Civile e Ambientale, Politecnico di Milano, Piazza L. Da Vinci 32,  Milano 20133, Italy}

\author{David Scheidweiler}
\affiliation{Institute of Earth Science, University of Lausanne, Lausanne 1015, Switzerland}
\affiliation{MRC Toxicology Unit, University of Cambridge; Tennis Court Road, Cambridge CB2 1QR, United Kingdom}

\author{Nolwenn Delouche}%
\affiliation{Institute of Earth Science, University of Lausanne, Lausanne 1015, Switzerland} 

\author{Alberto Guadagnini}%
\affiliation{Dipartimento di Ingegneria Civile e Ambientale, Politecnico di Milano, Piazza L. Da Vinci 32, Milano 20133, Italy}
\affiliation{Sonny Astani Department of Civil and Environmental Engineering, Viterbi School of Engineering, Los Angeles, California 90089-2531, USA}%

\author{Pietro de Anna}
\affiliation{Institute of Earth Science, University of Lausanne, Lausanne 1015, Switzerland}

\maketitle
%\linenumbers

\section*{abstract}
Biofilms alter the hydraulic properties of porous media, impacting processes from groundwater remediation to industrial filtration. While biomass accumulation is known to reduce permeability, a quantitative link between its spatial organization and system-scale hydraulics remains missing. Here, using microfluidics, time-lapse microscopy, and a novel mechanistic model we demonstrate that biofilm spatial organization is the key control for the resultant permeability decline. With independent experiments, we show that motile \emph{Pseudomonas putida} sp. and its non-motile mutant grow biofilm attaining identical total biomass, yet cause permeability reductions of 78\% and 94\%, respectively. This divergence arises because motile cells, escaping nutrient-depleted zones, colonizes the porous system differently in space, confining significant biomass upstream, whereas non-motile cells persistently colonize homogeneously the entire system. Our model, conceptualizing the medium as a series of pores with different size and biomass-modified permeability, accurately predicts these dynamics. We conclude that biomass spatial distribution, not simply its abundance, is the primary control on permeability, offering a new framework to predict and manage clogging in environmental and engineered systems.

\section*{Introduction}

\noindent
Porous media, such as soils, aquifers, and filters, serve as ideal habitats for sessile bacterial biofilms, which colonize the solid matrix surfaces while growing within the pore space~\cite{costerton1995microbial, conrad2018confined}. The role of biofilms is critical in various applications as they catalyze and facilitate bio-chemically driven processes underpinning engineering and environmental scenarios such as soil remediation~\cite{singh2021bacterial, cunningham2003subsurface}, bio-mineralization~\cite{ phoenix2008benefits, reith2006biomineralization}, design and formation of bio-barriers for containment of subsurface contamination~\cite{lennox2009biofilms}, water treatment~\cite{mahto2022bacterial}, or enhanced oil recovery~\cite{haddad2023investigating}. Biofilm dynamics can also lead to clogging in medical devices and industrial filtration systems~\cite{drescher2013biofilm, singh2021bacterial, schubert2002hydraulic, baveye1998environmental}, thus dramatically reducing their permeability, i.e., the ability to transmit fluid~\cite{coyte2017microbial}. Biofilms are complex structures where bacteria are embedded in a self-produced extracellular polymeric substance (EPS) matrix. The latter is primarily composed of lipids, proteins, exo-polysaccharides, and eDNA~\cite{flemming2016biofilms,davey2000microbial}. The EPS matrix serves multiple purposes, including providing mechanical stability and shielding microbial cells from environmental stresses, nutrient fluctuations, dehydration, antimicrobial agents, and shear forces. It also supports the retention of water and nutrients~\cite{flemming2010biofilm, flemming2016biofilms, wilking2011biofilms}. Additionally, the EPS matrix promotes intercellular communication by mediating the exchange of signaling molecules. This enables bacterial communities to coordinate their behavior within porous systems and adapt to changing environmental conditions~\cite{parsek2005sociomicrobiology, sionov2022targeting} even in the presence of a background fluid flow within porous systems~\cite{scheidweiler2024spatial}. \\
     
\noindent
The fluid movement is a key transport mechanism to convey resources across complex pore spaces. It critically influences biofilm development~\cite{coyte2017microbial}, affecting colony morphology~\cite{ceriotti2022morphology}, microbial interactions~\cite{wang2022dynamic, hartmann2019emergence, battin2016ecology, stoodley1998oscillation}, and cell motility~\cite{xavier2007cooperation,busscher2006microbial, klausen2003involvement, rodesney2017mechanosensing}. During their growth through individual cell division, biofilms progressively fill the pore space, altering its geometry and connectivity ~\cite{thullner2002influence, cunningham1991influence, karimifard2021modeling,aufrecht2019pore}. This process can lead to partial or complete occlusion of flow pathways for fluids, thereby increasing hydraulic resistance and significantly altering flow dynamics across the medium~\cite{drescher2013biofilm,coyte2017microbial}. These structural changes imprint onto the macroscopic properties of the porous medium, particularly porosity and permeability~\cite{kim2000biomass, gaol2021investigation, taylor1990substrate}, which are critical quantities shaping fluid flow and solute transport across the pore space~\cite{perez2022contributions,thullner2002influence}. As a consequence of the dynamic change in medium hydraulic properties, transport and availability of fresh resources is also affected, with important impacts on biofilm growth. \\

\noindent
The nature of the feedback between biofilm growth and flow conditions is complex, with biofilms both affecting and being affected by the hydrodynamic environment taking place across the pore space~\cite{purevdorj2004biofilm, taylor1990substrate, thullner2010comparison, gaol2021investigation,hartmann2019emergence}. Hydrodynamic conditions, such as shear stress, flow velocity, and pressure gradient, contribute to shape biofilm morphology, including its local thickness and density, as well as its spatial distribution~\cite{secchi2020effect,blauert2015time}. For example, high shear stresses can restrict biofilm thickness, resulting in a more compact structure. In contrast, low shear forces enable the development of thicker biofilms. When biomass clog pores, it decreases the medium permeability, and dramatically disrupt flow dynamics~\cite{paul2012effect, purevdorj2004biofilm}. The degree to which biofilms affect permeability is governed by various factors, including composition of the biofilm, its spatial distribution, and the inherent heterogeneity of the porous medium~\cite{kim2000biomass, kone2014impact, taylor1990substrate, cunningham1991influence}. Understanding this dynamic interplay between biofilm growth and hydrodynamics is critical for managing fluid flow and microbial processes in natural and engineered porous systems. Yet, the mechanisms governing the interaction between biofilm growth and the hydrodynamic environment are still poorly documented and understood, particularly with reference to the way biofilm spatial organization impacts macroscopic flow dynamics across porous media. \\

\noindent
This study investigates and documents how biomass growth controls permeability of porous structures where fluid flow is driven by a constant pressure gradient. This scenario closely mirrors realistic natural settings, such as wet-soil environments, where hydraulic gradients (pressure drops) induce the fluids flow and, eventual, medium permeability dynamics alters flow and resources transport, while in other settings, e.g. industrial, pumps-imposed flow generates macroscopic hydraulic gradients that can change dynamically together with the host medium permeability resulting in biomass removal~\cite{kurz2022competition}. We designed and engineered a customized pressure control system that enables us to maintain a constant macroscopic pressure gradient by fully incorporating the effects of pressure loss in the connecting pipes (that depend on local flow conditions) and variations of fluid volume in the inlet and outlet reservoirs~\cite{jiao2024intrinsic}. While the pressure gradient across the porous medium remains fixed, fluid flow rate is continuously monitored using an analytical scale, thus enabling accurate assessment of the (macroscopic) permeability of the medium. We focus on the growth patterns of motile (flagellated) and non-motile (non-flagellated mutant strain) soil bacteria \textit{Pseudomonas putida} sp. inside a microfluidic circuit specifically designed to mimic a complex pore geometry, typical of several systems (such as natural soils or engineered industrial filters). The motility trait, allowing bacteria to actively navigate porous media and respond to environmental conditions, leads to distinct spatial distributions of the biomass compared to non-motile strains. While the overall biomass of the two strains attains the same carrying capacity, we document a significantly different reduction of about $78 \pm 7 \%$ and $94 \pm 4 \%$ of the initial overall permeability for motile and non-motile, respectively. We provide an interpretation of this remarkable result through a model that we specifically develop to embed the effect of spatial biomass growth on permeability changes. \\

\begin{figure*}[htb!]
\centering
\includegraphics[width=1\linewidth]{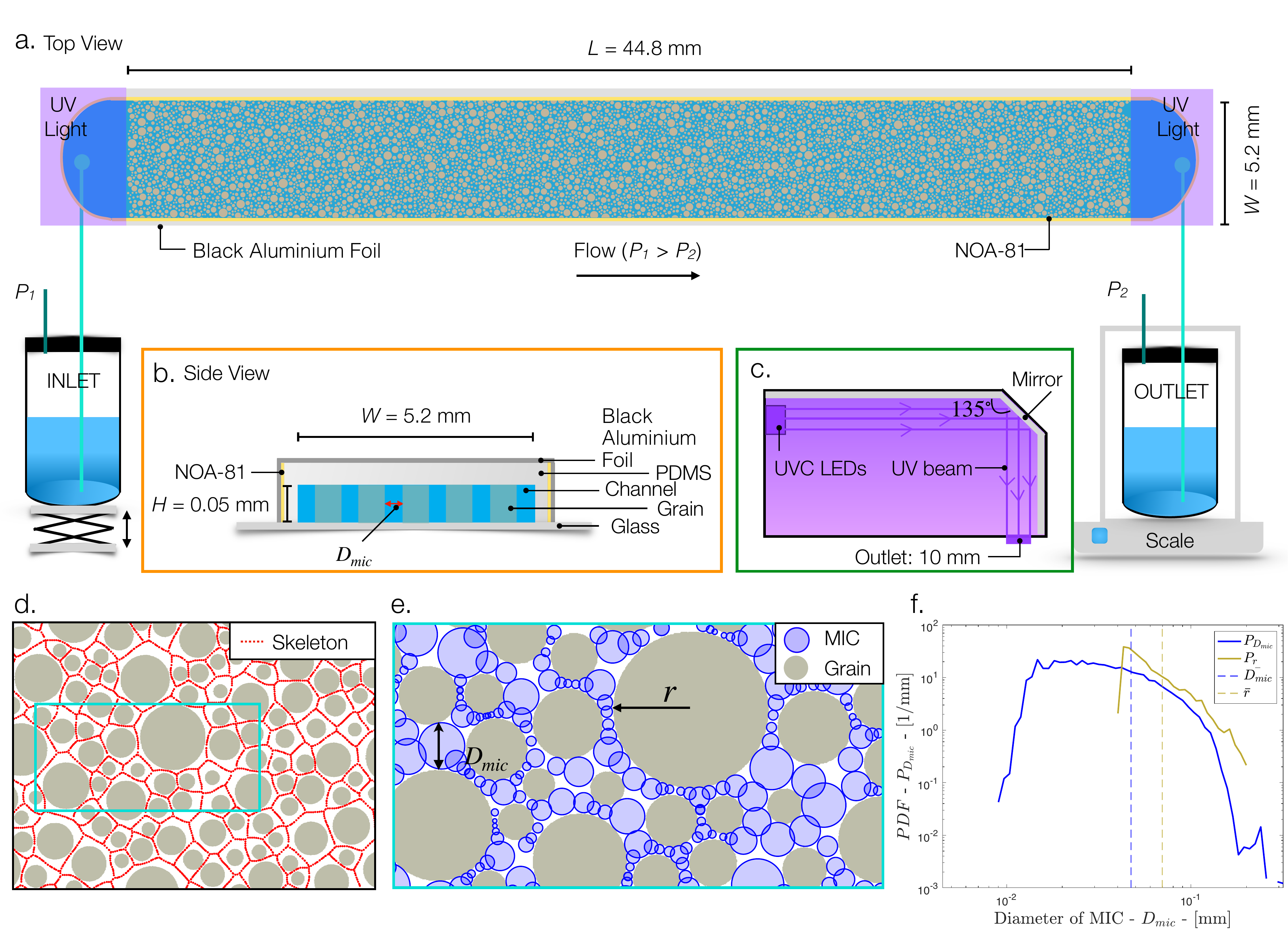}
\caption{Microfluidic system designed for monitoring intrinsic permeability and controlled flow dynamics. (a) Schematic of the experimental setup used to quantify the intrinsic permeability of the host porous medium. A constant pressure difference ($P_1 - P_2$, with $P_1 > P_2$) is applied across the microfluidic device while the outlet reservoir is continuously weighed to monitor flow rate. The porous medium is designed composing disks (gray) to mimic the solid matrix. Falcon tubes are used as inlet and outlet reservoirs, sealed with microfluidic adapters (black caps). (b) Cross-section*al view of the microfluidic device: A PDMS channel, (blue region) with embedded pillars (gray) of the porous structure, is plasma-bonded to a glass slide. The pore space between pillars is shown in blue. The PDMS, with the exception of the top surface, is coated with gas-impermeable NOA-81 (yellow). (c) UV-C device schematic: A 3D-printed guide with mirrors (gray) reflects UV-C light into specific regions of the chip, preventing unwanted bacterial colonization in the inlet and outlet zones. (d) Schematic illustration of a portion of the porous medium skeleton (red dashed lines), generated from the binarized image to capture connectivity of the pore space. (e) Results of a Maximum Inscribed Circle (MIC) algorithm applied to the pore structure (blue circles represent the largest inscribed circle within each pore). The centers of these circles are aligned with the skeleton (red dashed lines) and positioned where they touch the closest grain. (f) Double logarithmic plot of the probability density function (PDFs) of grain size \( P_r \) (gray) and pore (MIC) diameter \( P_{D_{\text{mic}}} \) (blue). The dashed lines indicates the average pore size (blue) \( \bar{D}_{\text{mic}} = 47 \, \mu m \) and grain diameter (yellow) \( \bar{r} = 70 \, \mu m \). To prevent redundancy, any pair of adjacent MICs with an overlap area exceeding 15\% of the sum of their areas is processed by randomly removing one of the them.}
\label{Fig1}
\end{figure*}

\section*{Materials and Methods}

\subsection*{Porous structure and microfluidics fabrication}

\noindent
We use microfluidics devices as porous media analogs to mimic natural complexity of a subsurface environment. The chip ($L=44.8$~mm $w = 5.2$~mm) contains a random distribution of vertical cylinders (grains) with distributed radius $r$ ranging between $40$ and $200 \, \mu$m (see Fig.~\ref{Fig1}~$a$). The thickness of the device, $H = 0.05$~mm, is comparable to the average pore opening, $\overline{\lambda} = 0.07$~mm (representative of many common porous structures), see Fig.~\ref{Fig1}~$b$. This design yields parabolic-like fluid velocity profiles in both vertical and horizontal directions between solid boundaries, and results in a heterogeneous velocity distribution at the mid-depth~\cite{de2021chemotaxis,bordoloi2022structure}. 

\noindent
The resulting porous system contains an array of cylindrical PDMS pillars, which represent the solid matrix of the porous system. The skeleton of the designed geometry is shown as a red solid line in Fig.~\ref{Fig1}~$d$: it represents the pore space locations equally distant from the two closest grain walls. We evaluate the Maximum Inscribed Circles (MIC) along the skeleton representing the local pores, the diameter of each MIC denoting the size of the local pore (Fig.~\ref{Fig1}~$e$)~\cite{bordoloi2022structure}. The measured average pore size is $\bar{D_{mic}} = 47\,\mu$m, lies within a typical range of value observed in natural soil environments \cite{bear1988dynamics, taylor2004implications}. The overall pore size distribution is shown in Fig.~\ref{Fig1}~$f$. \\

\noindent
We aim at observing the growth of bacterial cells in a porous systems where the resources available for cell division (mostly nutrients and oxygen) are associated with the incoming fluid flow injection. Thus, to limit non-uniform oxygen permeation through the PDMS, a UV-curable, gas-impermeable resin (Norland Optical Adhesive, NOA-81 Norland Products Inc~\cite{ceriotti2022morphology}) was applied to the lateral surfaces of the device. The top surface was intentionally left uncoated to preserve optical transparency for imaging. This configuration provides a spatially uniform oxygen supply, whose diffusion occurs through the top PDMS layer across the whole porous domain. The NOA-81 was then cured using an LED light source (M365LP1, \textit{Thorlabs}) with a nominal wavelength of 365~nm~\cite{ceriotti2022morphology}. Finally, to eliminate any air bubbles trapped during the initial chip saturation, the microfluidic device was degassed in a desiccator for approximately 30 min before the first fluid injection \cite{scheidweiler2020combining}.\\

\subsection*{Bacterial strains and growth condition}

\noindent
Flow experiments of biofilm growth were performed using \textit{Pseudomonas putida} sp. and its derivative: the wild-type (WT) \textit{P. putida} KT2440 strain (tagged to express green fluorescent protein, GFP) and the non-flagellated mutant \textit{P. putida} KT2440 $\Delta$filC strain (also tagged with GFP)~\cite{arce2016rna}. The $\Delta$filC mutant, lacking flagella due to a deletion of the \textit{filC} gene, served as the non-motile variant, while the WT strain retained normal flagellar function and, hence, motility. Both strains are resistant to gentamicin. \textit{P. putida} KT2440, a non-pathogenic soil-dwelling bacterium, is well-known for its biofilm formation capabilities and metabolic versatility, making it a potential agent for bio-remediation applications in industrial and environmental contexts \cite{nelson2002complete, belda2016revisited, sathesh2021inducible}. \\

\noindent
For the preparation of bacterial cultures, frozen stocks of \textit{P. putida} KT2440 $\Delta$filC GFP-tagged and \textit{P. putida} KT2440 WT GFP-tagged were separately inoculated into 4~mL of sterile Luria-Bertani (LB) broth (25~g/L) supplemented with 4~$\mu$L of gentamicin (15~mg/L). The cultures were incubated overnight at 30 \textdegree{}C, shaking at 180 rpm. Thus, 40~$\mu$L of each overnight culture was diluted 1:100 into 4~mL of fresh LB broth at 30\% concentration with 4~$\mu$L of gentamicin (15 mg/L) and further incubated at 30 \textdegree{}C with shaking at 180 rpm for 4~h, so that early exponential phase was reached \citep{scheidweiler2020trait}. The prepared bacterial suspensions were used to infect the microfluidic chip, injecting them with a micro-pipette through the outlet until the chip was saturated. After inoculation, the bacterial suspension was allowed to adhere and adapt to the porous structure under static (no-flow) conditions for 30 minutes before initiating sterile nutrient flow  driven by a macroscopic pressure drop. This adaptation period, consistent with prior studies ~\cite{scheidweiler2024spatial, scheidweiler2020trait}, promotes initial cell retention while minimizing early-stage biofilm development. Varying this duration is expected to mainly shift the temporal onset of biomass growth and permeability evolution, without qualitatively affecting the underlying dynamics. \\

\subsection*{Pressure control system}

\noindent
We developed a microfluidic pressure control system, similar to~\cite{jiao2024intrinsic}, to precisely regulate and maintain a stable pressure drop, \( \Delta P_{\text{chip}} \), between the inlet and outlet of the microfluidic device, thereby ensuring reliable and repeatable experimental conditions. The imposed pressure boundary conditions mimic those expected in natural soils or in several filtration systems, where flow is driven by gravity. Designing this system was key to ensure stable macroscopic pressure conditions for biofilm growth within the microfluidic chip throughout the experiment. In fact, we want here to avoid an imposed flow rate (as with a syringe pump) that, with biomass growth, would result in local pressure increase and biomass removal. The setup integrated a pressure controller (OBI1 - MK3, \textit{ElveFlow}) and an analytical scale (XS205DU, \textit{Mettler-Toledo}) to generate a pressure-driven fluid motion and to continuously monitor the associated macroscopic flow rate, \( Q(t) = \frac{\Delta M(t)}{\rho \Delta t} \), where \( M(t) \) represents the mass of the outlet reservoir at each time $t$, \( \rho \) is the density of the flowing solution and $\Delta t$ the time difference between two measurements. In the Supplementary information this system is described in detail.

\subsection*{UV-C dose to limit biomass growth outside the porous channel}

\noindent
Following the work~\cite{ramos2023ultraviolet}, a source of ultraviolet-C (UV-C) light diodes (LEDs) (1W-20mm-120\textdegree{}C, \textit{TiaoChongYi}), emitting at a wavelength of 270~nm, was employed to spatially confine bacterial growth to the pore space, as illustrated in Fig.~\ref{Fig1}~$a$. To prevent unintended biomass growth outside the porous domain (which would affect flow rate and thus permeability measurements) UV-C illumination was applied to the inlet/outlet regions and connecting tubing. The porous section* of the microfluidic device was shielded with black aluminum foil (Fig.~1a) to prevent UV exposure within the pore space. In absence of the UV-C treatment, biomass also grows upstream of the porous system, periodically releasing cell aggregates into the flowing fluid. These aggregates are transported into the system, where they accumulate and alter both biofilm development and the resulting macroscopic permeability. This approach is highly effective in (confining) the growth of \textit{P. putida} KT2440 cells to the desired regions~\cite{ramos2023ultraviolet}.

\subsection*{Time-lapse video-microscopy} 

\noindent
Time-lapse imaging was performed through an inverted and fully automated microscope (Eclipse Ti2, Nikon) equipped with a sCMOS camera (Hamamatsu ORCA flash 4.0, 16-bit) and controlled by Nikon Elements software. This integrated system allows for automatic capture of large images time series. A fluorescence optical configuration was used for capturing the time series images, with each individual picture acquired at $M = 4X$ magnification, corresponding to a spatial resolution given by the camera sensor pixel size divided by $M$: $6.5 \, \mu$m$/4 = 1.625\, \mu$m/pixel. All individual images were captured at the camera full resolution of $2048  \times  2048$ pixels. The acquisition system covered the entire porous domain of the microfluidic device by stitching $19 \times 3$ individual images into a composite array. For fluorescence imaging, we used a Nikon GFP-HQ filter with a Spectra X-light engine to excite the fluorescence signal from the GFP-tagged \textit{P. putida} cells~\cite{scheidweiler2024spatial}.

\subsection*{Biomass growth}

\noindent
The biomass accumulation over time were quantified through time-lapse images (see Supplementary Information for more details). The primary metric used was $B_{\text{exp}}$, which represents the biomass in terms of pixel count, normalized by the total area of the porous part of the microfluidic chip. Fluorescence intensity is used here as a proxy for biomass distribution and accumulation. This approach assumes an a linear relationship between fluorescence signal and biomass over the range considered, this assumption being potentially affected by signal saturation, optical attenuation, and structural effects such as cell stacking or EPS shielding. Accordingly, fluorescence should be interpreted as providing a relative, rather than absolute, biomass estimate. The classical Logistic Growth (LG) model~\cite{scheidweiler2019unraveling} was calibrated to the experimental observations, to describe temporal biofilm accumulation ($B_{\text{t}}(t)$). After the initial exponential growth, characteristic of the cell division mechanism, the biomass growth slows down and stabilizes about the carrying capacity. The macroscopic LG model is characterized by three parameters: initial biomass ($P_0$), carrying capacity ($K$), and growth rate ($G_r$) as:
\begin{equation}
    B_{\text{t}}(t) = \frac{K \cdot P_0 \cdot \exp(G_r \cdot t)}{K + P_0 \cdot (\exp(G_r \cdot t) - 1)}.
\end{equation}
The doubling time, $t_{\text{d}}$, corresponds to the time required for the biomass to double in size (can be interpreted as the average time needed by a cell to divide), and it is given by $t_{\text{d}} = \frac{\ln(2)}{G_r}$.\\

\noindent
We also monitored the biomass at microscopic level, within pores. We measured the biomass density within each pore. For each pore, denoted with $i$, the intensity-based biomass density, $\rho_{\text{pi}_i}$, is defined as the total pixel intensity within the pore, normalized by the maximum intensity difference between the final and initial images, and by the area $A_{\text{mic}_i} = \pi D_{\text{mic}_i}^2 / 4$ of that pore.

\subsection*{Permeability dynamics measurement}

\noindent
As biomass develops within pores, it obstructs fluid flow. This yields a progressive reduction in permeability. To quantify this process, we assess the time-dependent permeability \( k_{exp}(t) \) upon relying on Darcy’s Law~\cite{darcy1856fontaines, bear1988dynamics}:\\
\begin{equation}
    q(t) = - \frac{k_{exp}(t)}{\mu} \, \nabla P_c(t),
\end{equation}
where \( q(t) = \frac{Q(t)}{WH}\) is the volumetric flow rate at time \( t \), and \( \nabla P_c(t) = \frac{\Delta P_c(t)}{L}\) is the pressure gradient across the grain-containing section* of the porous medium, which is comprised between the (grain-free)inlet and outlet zones.  \\

\noindent
Since, inlet area, porous part and outlet area are designed in series, the total pressure drop across the porous domain, \( \Delta P_c(t) \), is determined as the overall chip pressure drop, subtracting the pressure losses in both the grain-free section*s of the chip \( \Delta P_{l_2}(t) \), inlet and outlet, according to:
\begin{equation}
   \Delta P_c(t) = \Delta P_{\text{chip}}(t) - \Delta P_{l_2}(t),
\end{equation}
\noindent
where \( \Delta P_{l_2}(t) \) represents the pressure losses due to fluid flow in the grain-free section*s of the chip. This pressure loss is calculated from Stokes flow solution through a pipe of rectangular section* of width $W$ much larger than its thickness $H$, as follows~\cite{bruus2007theoretical}:
\begin{equation}
    \Delta P_{l_2}(t) = -\frac{12 \mu Q(t) L_{gf}}{H^3 W \left(1 - \sum_{\eta=0}^{\infty} \frac{192}{(2\eta +1)^5 \pi^5} \frac{H}{W} \tanh\left(\frac{(2\eta +1)\pi W}{2H}\right)\right)},
\end{equation}
\noindent
where \( L_{gf} \) is the length of the grain-free section* of the chip, \( \eta \) is a positive integer representing the higher-order corrections for flow in the rectangular channel. By isolating the pressure drop \( \Delta P_c(t) \) across the grain-containing section* of the chip, the time-dependent permeability \( k_{exp}(t) \) can be determined as the biofilm develops and progressively clogs the pore spaces.

\subsection*{Flow experiment}

\noindent
 Following the inoculation, we activated the pressure control system to ensure a stable pressure differential across the microfluidic chip (\( P_{\text{chip}} \)). Initially, the inlet pressure \( P_1 \) was set to 16~mbar, while the outlet pressure \( P_2 \) was maintained at 10~mbar. Pressures \( P_1 \) and \( P_2 \) correspond to the reservoir pressures at the inlet and outlet of the system, respectively. These are dynamically varied to maintain a constant pressure gradient across the porous part of the entire device. The inlet reservoir was filled with 30\% LB medium with gentamicin (15 mg/L) to avoid contamination (the used strains are resistant to gentamicin), while the outlet reservoir initially contained Milli-Q water. The outlet reservoir was continuously weighed using an analytical balance to monitor flow rates (evaluated from the mass of the fluid in the reservoir over time). \\

\noindent
All experiments were performed at a constant temperature (30 \textdegree{}C) via a microscope incubator (OKOlab) and conducted in triplicate to ensure reproducibility. As the sterile nutrient solution start flowing, the experiment begins and we start collecting time-lapse large images to monitor biomass growth and recording the outlet reservoir mass. Images were captured every 30 minutes for the first 3~h, followed by intervals of 1~h up to 48~h.

\section*{Results}

\subsection*{Porous structure characterization}
\noindent
We characterize the heterogeneous geometry of the pore space by computing along the structure skeleton (see red paths in Fig.~\ref{Fig1}~$d$) the Maximum Inscribed Circle (MIC; identifying each pore as a circle, see blue disks in Fig.~\ref{Fig1}~$e$). The diameter of the latter ($D_{MIC}$) denotes the distance between grain walls and is typically employed as a representative of the pore size. As depicted in Fig.~\ref{Fig1}~$f$ (blue curve), it ranges between $0.01$ and $0.2$~mm across our system. Intrinsic permeability associated with this complex structure could be assessed through a mechanistic model~\cite{jiao2024intrinsic} according to which flow through the whole system is assumed to be described upon representing the domain as a sequence of small porous systems in series, each associated with a permeability rendered by the Hagen-Poiseuille law~\cite{bear1972dynamics}. This results in an intrinsic permeability of about 50~darcy, as verified by direct evaluations (based on the use of Darcy’s law) of the overall system permeability. The fabricated chip is then inoculated with a bacterial suspension and its permeability is monitored as bacteria divide and biomass grows under flow of nutrients driven by the imposed macroscopic and constant pressure drop.

\subsection*{Porous medium colonization by \textit{Pseudomonas putida}}

\noindent
We prepared bacterial cultures of motile and non-motile bacteria: \textit{Pseudomonas putida} sp. KT2440 wild-type (WT) modified to express green fluorescent protein GFP and its non-flagellated mutant (\( \Delta \)fliC) also tagged with GFP (see S.I.). Immediately after chip inoculation, the UV-C LEDs (Fig.~\ref{Fig1}~(c)) were activated to continuously irradiate the grain-free regions of the chip. After 30 minutes since inoculation, we switch to the continuous injection of a sterile LB solution. Bacteria can divide, and the associated biomass grows only within the porous landscape, as a continuous UV-C illumination is guaranteed at the inlet and outlet for the whole duration of the experiment. In both scenarios (corresponding to WT and $\Delta$fliC), biomass growth is driven by division of cells that are attached to the solid surfaces of the chip while up-taking nutrients and respiring oxygen transported by the flow. The consequent biomass growth and accumulation in the pores reduces the space available for fluid to flow. Pores are eventually filled and (in some cases) completely clogged. When the latter condition takes place, flow within a given pore is dramatically reduced. As a biofilm is indeed a porous material itself, the flow is never expected to completely stop as the fluid can pass among cells and within the EPS~\cite{billings2015material, flemming2000biofilms, pintelon2012effect, dreszer2013hydraulic, mcdonogh1994permeability}. \\

\begin{figure}[htb!]
    \centering
    \includegraphics[width=1\linewidth]{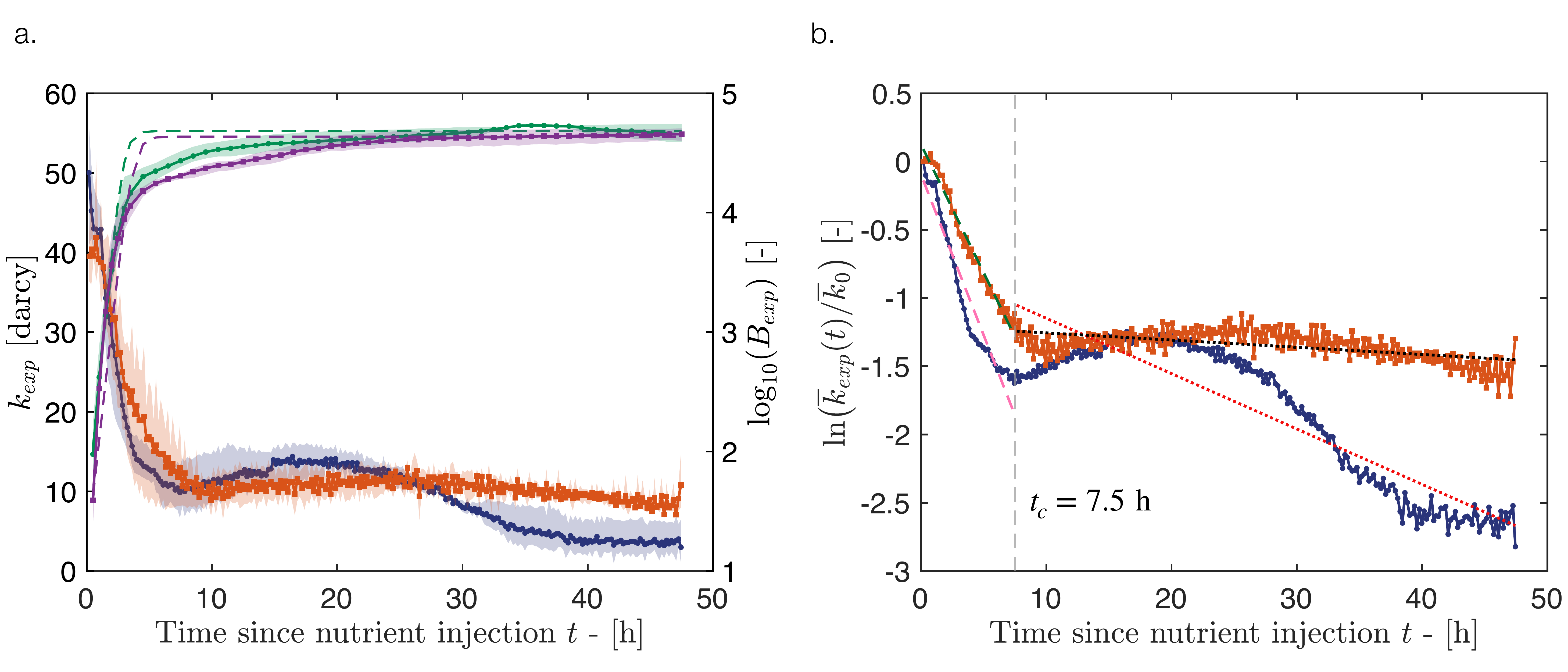}
    \caption{%
    Permeability, biomass, and clogging kinetics for $\Delta \mathrm{fliC}$ and WT strains. \textbf{(a)} Dynamics of the experimental permeability ($k_{exp}$) and biomass ($B_{exp}$) since sterile nutrient injection. The average among three replicas of the experimental permeability is shown as blue ($\Delta \mathrm{fliC}$) and orange (WT) dots (shaded area represents standard deviation among replicas), while the average experimental biomass is shown in green ($\Delta \mathrm{fliC}$) and purple (WT) dots (shaded areas begin standard deviation). The experimental biomass ($B_{exp}$) is inferred from the GFP signal emitted by the cells over time. Dashed light-green and light-purple lines denote the fitted Logistic Growth Model prediction (Methods.F; denoted as $B_{\text{t}}$) for the $\Delta \mathrm{fliC}$ and WT strains, respectively. \textbf{(b)} Assessment of permeability dynamics, represented as $\ln\!\left(\overline{k}_{exp}(t)/\overline{k}_0\right)$, where $\overline{k}_{exp}(t)$ denotes the strain-averaged permeability and $\overline{k}_0$ its initial value. Blue and orange curves correspond to the $\Delta \mathrm{fliC}$ and WT strains, respectively (same data as in $a.$). Dashed pink and green lines indicate the early-times linear fit for the $\Delta \mathrm{fliC}$ and WT strains, respectively, while the dotted red and black lines indicate the corresponding late-times linear fits. A transition time $t_c = 7.5$~h marks the early rapid-clogging regime from a later-times dynamics. Slopes of the piecewise linear fits in the figure provide an effective estimate of $d(\ln k)/dt$ and of the associated characteristic clogging timescales.}
    \label{Fig2}
\end{figure}

\noindent
The growth curve average between triplicates is shown in the semi-logarithmic plot of Fig.~\ref{Fig2}~$a$ (the shaded area corresponds to the standard deviation among replicas).  The growth curve (biomass detected as GFP signal emitted by the cells) increases exponentially fast for the first 7.5 ~hours in all experiments with motile (WT) and non-motile ($\Delta$fliC) strains. The exponential growth is, then, followed by a stationary phase during which the overall biomass stabilizes at a constant value $K$, corresponding to the biomass carrying capacity. Growth curves for WT and $\Delta$fliC strains display a similar behavior and are well captured by a logistic model: the measured (fitted) carrying capacity is $K = 4.6 \times 10^4$ for WT, and $K = 4.7 \times 10^4$ and for $\Delta$fliC, which means, substantially, the same overall biomass) 

\subsection*{Permeability drop due to biomass growth}

\noindent
Remarkably, although the overall biomass accumulation is nearly identical for the WT and $\Delta$fliC settings, the associated permeability reduction due to biomass colonization differs markedly. Figure~\ref{Fig2}~shows the permeability dynamics (solid blue curve and solid orange curve, averaged over three replicas; shaded area corresponds to standard deviation). For the $\Delta$fliC (non-motile) setting, figure~\ref{Fig2}~$a$ (solid blue curve) documents that, starting from the initial value of $k_0 = 50$~darcy, permeability initially decreases to about $10$~darcy, it then remains almost constant, to finally decrease, after 25 hours, to the $k = 3$~darcy (6\% of $k_0$, i.e. 94\% reduction) even though the overall biomass does not increases further after 25 hours. Otherwise, for the WT scenario, permeability reduces to $k = 10$~darcy (22~\% of $k_0$, i.e. 78\% reduction) to, then, remain stable. These observations suggest that biomass growth alone cannot account for the observed permeability decline.\\

\noindent
To quantify clogging kinetics, beyond the final permeability reduction, we analyzed the strain-averaged permeability dynamics in terms of the temporal evolution of logarithmic transformed permeability
\[
\ln\!\left(\overline{k}_{exp}(t)/\overline{k}_0\right),
\]
where $\overline{k}_{exp}(t)$ is the strain-averaged experimental permeability and $\overline{k}_0$ its initial value. For all investigated cases, we observe a characteristic transition time $t_c = 7.5$~hours (see Fig.~2b) to distinguish an early rapid-clogging regime from a later-stage dynamics. Within selected temporal windows demarcated by $t_c$, the evolution of the above quantity is fitted by a linear model. The associated measured slope represents a temporal derivative and defines a characteristic clogging timescale $\tau = -1/\mathrm{slope}$. For the $\Delta \mathrm{fliC}$ strain, the early- and late-stage slopes are $-0.24$~h$^{-1}$ and $-0.04$~h$^{-1}$ respectively. These correspond to characteristic timescales of $4.25$~hours and $24.66$~hours, respectively. The early- and late-stage slopes for the WT strain are $-0.19$~h$^{-1}$ (similar to the $\Delta \mathrm{fliC}$ strain) and $-0.005$~h$^{-1}$ (markedly different from to the $\Delta \mathrm{fliC}$ strain), respectively. These fitted slopes yield characteristic timescales of $5.32$~hours and $190$~hours, respectively, suggesting that both strains undergo an initial rapid permeability reduction. However, the later-stage evolution differs substantially, the $\Delta \mathrm{fliC}$ strain continuing to exhibit slow clogging, whereas the WT strain transitions to a much weaker, almost absent, permeability dynamics. The results of this analysis are shown in Fig.~2b.\\
\begin{figure}[htb!]
\centering
\includegraphics[width=1\linewidth]{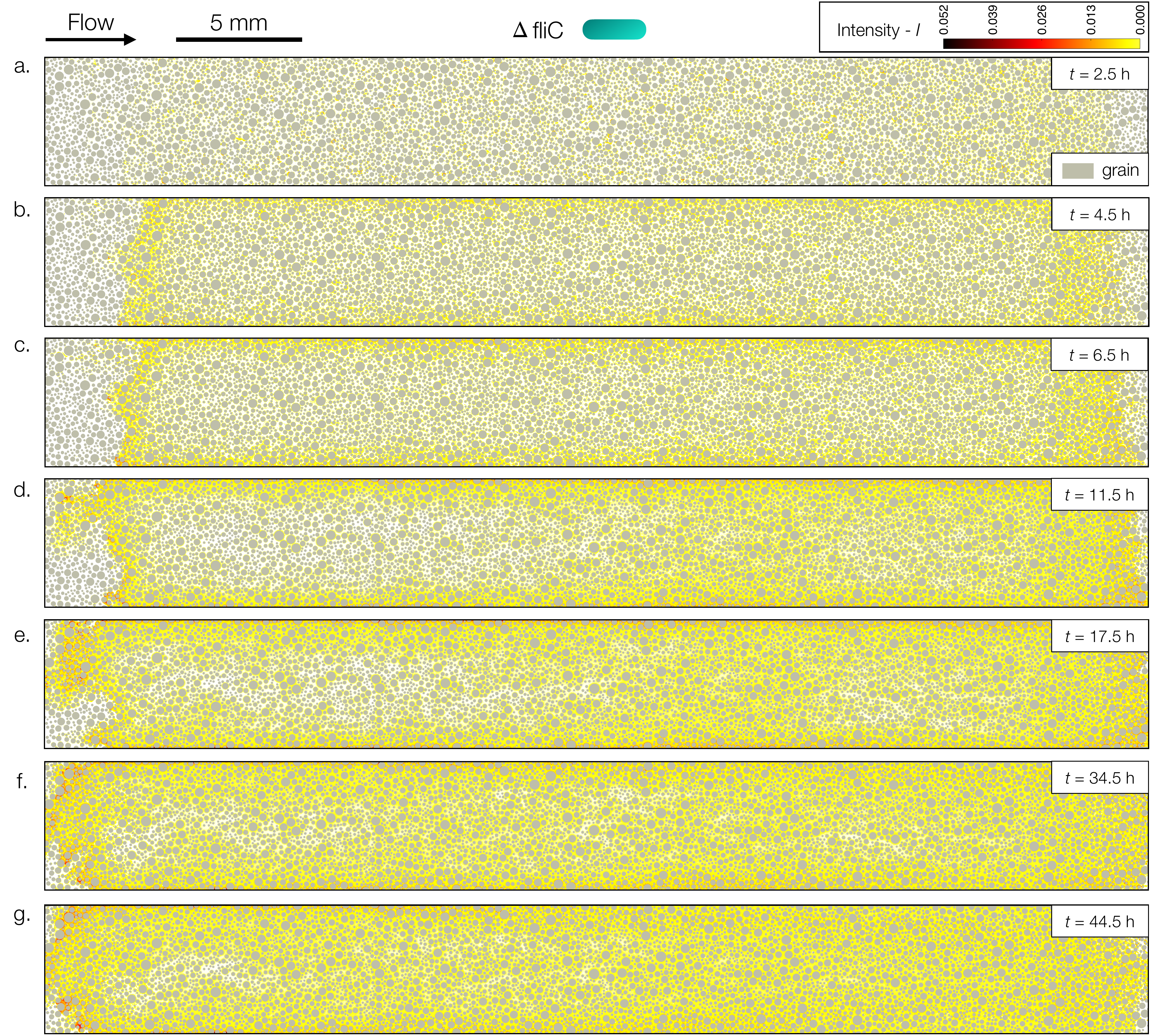}
\caption{Dynamics of \textit{P. putida} sp. $\Delta$fliC biomass as GFP signal (increasing light intensity going from light to dark) growing between the solid grains (gray disks) observed at various times ($t = 2.5$ h, $t = 4.5$ h, $t = 6.5$ h, $t = 11.5$ h, $t = 17.5$ h, $t = 34.5$ h, and $t = 44.5$ h).}
\label{Fig3}
\end{figure}

\begin{figure}[htb!]
    \centering
    \includegraphics[width=1\linewidth]{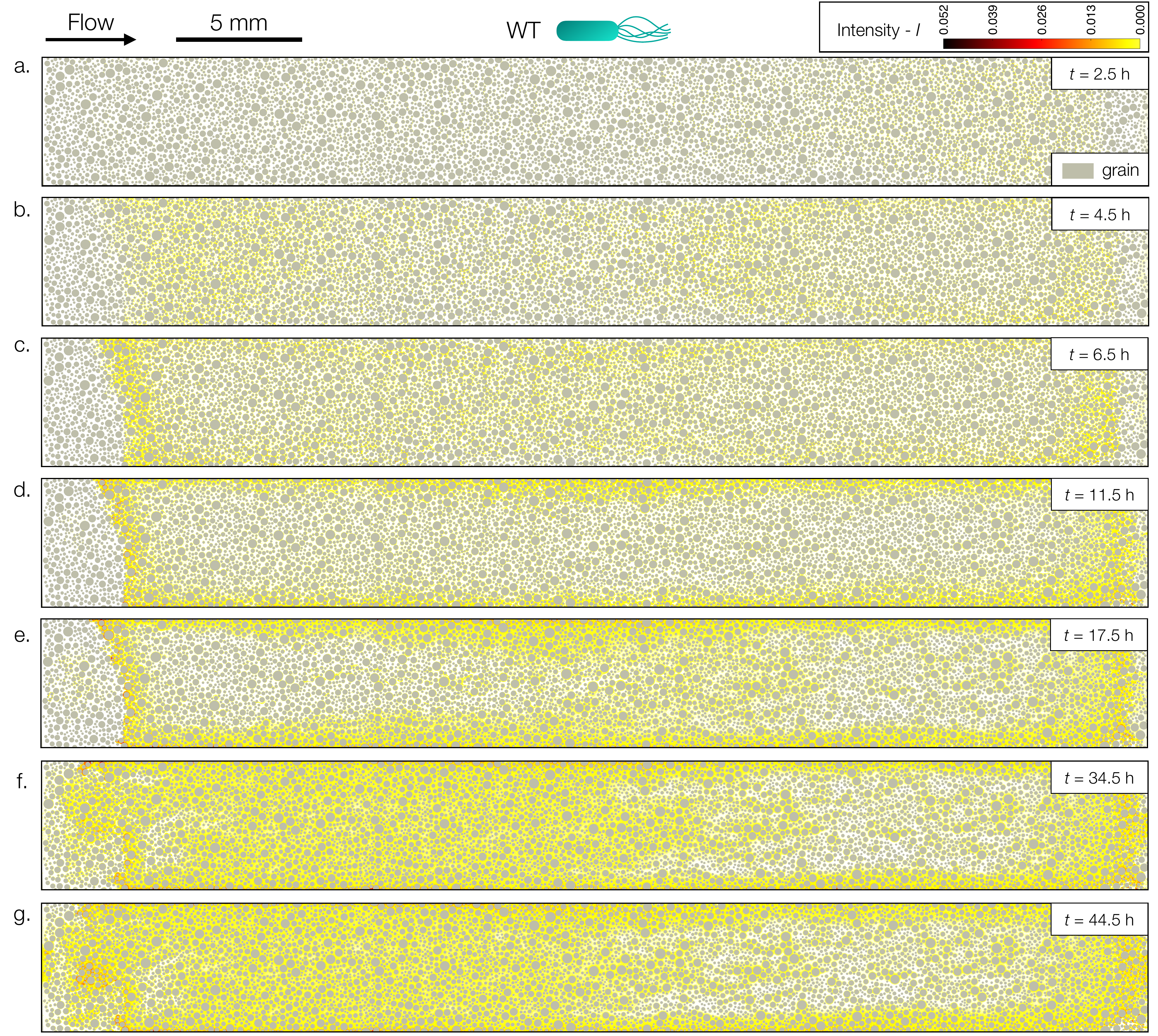}
    \caption{Dynamics of \textit{P. putida} sp. WT biomass  as GFP signal (increasing light intensity going from light to dark) growing between the solid grains (gray disks) observed at various times ($t = 2.5$ h, $t = 4.5$ h, $t = 6.5$ h, $t = 11.5$ h, $t = 17.5$ h, $t = 34.5$ h, and $t = 44.5$ h).}
    \label{Fig4}
\end{figure}

\subsection*{Spatial organization of biomass growth}

\noindent
As shown in Fig.~\ref{Fig3} and Fig.~\ref{Fig4}, the spatial organization and dynamics of \textit{P. putida} sp. biomass for WT and $\Delta$flic, respectively, are quantified by the fluorescent light intensity detected (increasing light intensity going from light to dark) within the pore space (solid grains are gray disks in the figure). As discussed, biomass grows similarly (exponential grow; see also Fig.~\ref{Fig2}~$a, b$) during the initial 7.5 hours for both scenarios. Most of the bacterial growth takes place near the system inlet, where resources (nutrients and oxygen), transported by the incoming flow, are in high concentrations. At later times, between about 8 and 20 hours, the biomass growth slows down to attain a plateau, corresponding to a stationary phase (see Fig.~\ref{Fig2}~$a, b$). This corresponds to biomass accumulation also downstream where cells are exposed to reduced availability of resources (nutrients and oxygen), that are depleted upstream. As time progresses (after 25 hours), we observe a different behavior between the WT and $\Delta$fliC strains. In the second half of the porous domain (beyond ~20 mm, corresponding to about 400 average pore sizes), WT cells exhibit limited accumulation. We argue that they rely on motility to escape the resources-depleted region, swimming away and being advected downstream (outside the field of view) by fluid flow. In contrast, the non-motile strain $\Delta$fliC continues to growth with the limited resources available, slowly occupying the pore space homogeneously throughout the entire system. Spatial maps of biomass for additional replicates (illustrated in the Supplementary Information) confirm this observed behavior.

\section*{Discussion}

\subsection*{Model for porous media permeability with growing biomass}

\noindent
We develop a mechanistic modeling framework to quantify how biomass growth alters the medium permeability, bridging pore-scale structure and system-scale hydraulic response. The model operates as a hydraulic upscaling approach based on pore-scale biomass distributions, linking microscopic biomass spatial organization to macroscopic permeability. Nutrient and oxygen transport are not explicitly resolved; instead, their effects are implicitly captured through the experimentally observed biomass organization. The porous medium is represented as a series of elementary units aligned with the flow direction, each idealized as a pipe whose effective diameter is determined by the local pore geometry. Biomass growth is incorporated modifying the permeability of each pipe to account for reduction in effective pore size, which is caused by biomass accumulation at the pore (pipe) walls. Extensions of the model could incorporate explicit resource transport and growth dynamics to achieve fully predictive capability. \\

\begin{figure*}[htb!]
\centering
    \includegraphics[width=1\linewidth]{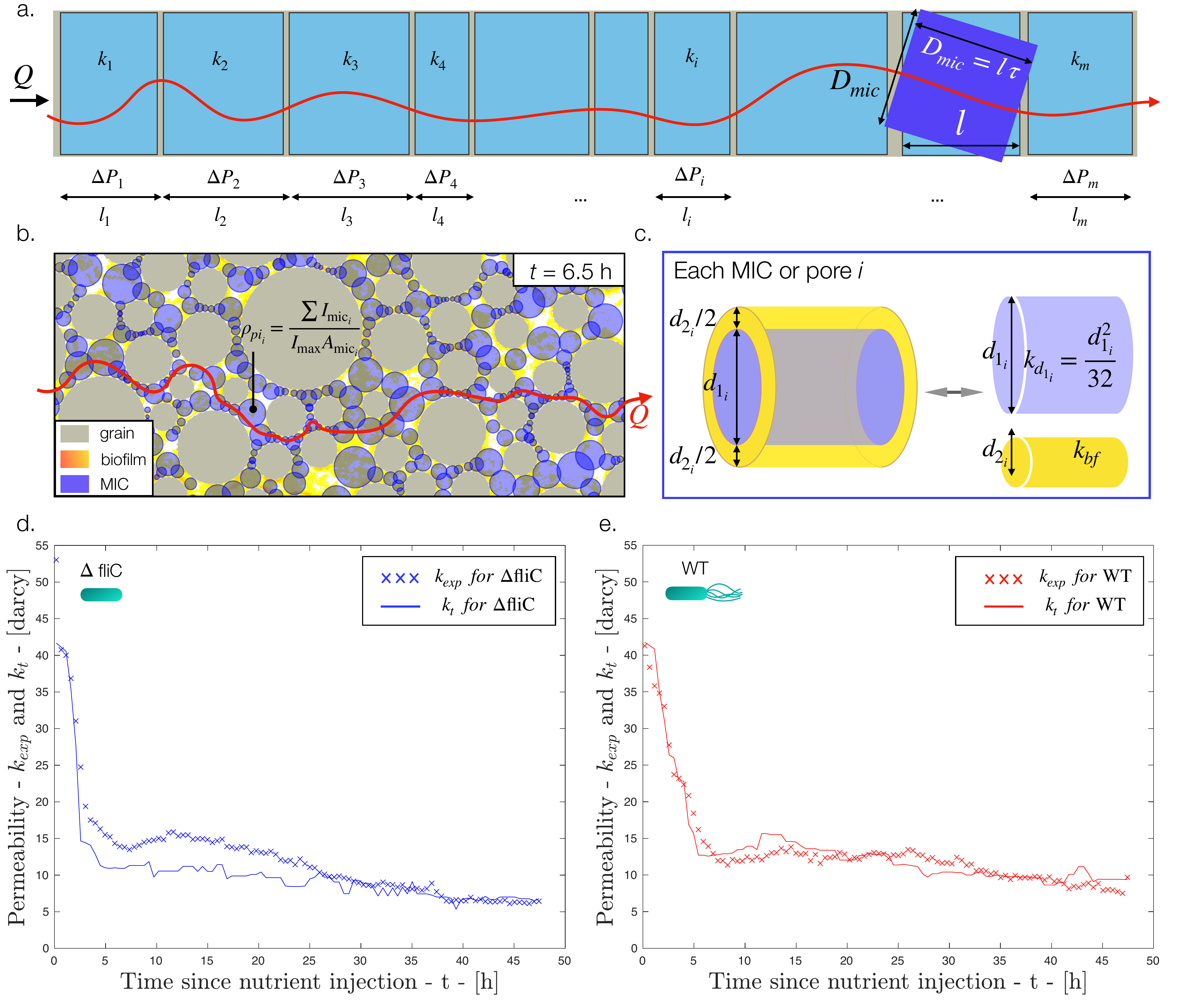}
    \caption{Theoretical model and predictions. (a) Schematic view of the permeability model, conceptualizing the system as a series of (virtual) porous media with individual lengths $l_i$ and permeability $k_i$~\cite{jiao2024intrinsic}. (b) biomass density $ \rho_{{pi}_i} = \frac{\sum I_{{\mathrm{mic}}_i}}{I_{{\mathrm{max}}} A_{{\mathrm{mic}}_i}}$ (in yellow) evaluated from fluorescence images, where $\sum I_{\mathrm{mic}_i}$ is the sum of pixel intensity within the pore \(i\), $I_{\mathrm{max}}$ denotes the maximum intensity contrast, and $A_{\mathrm{mic}_i}$ is the pore area. (c) Schematic of an individual pore with biofilm. Each pore \(i\) is represented as two parallel flow systems: (1) a central biofilm-free pipe with diameter \(d_{1_i}\) and permeability \(k_{\mathrm{d}_{1_i}} = d_{1_i}^2/32\); and (2) an annular biofilm-occupied region with thickness \(d_{2_i}/2\) and permeability \(k_{\mathrm{bf}}\). The equivalent local permeability is, then, given by their mean, weighted by their own thickness as \(k_{e_i} = \frac{d_{1_i} k_{\mathrm{d}_{1_i}} + d_{2_i} k_{\mathrm{bf}}}{d_{1_i} + d_{2_i}}\). (d, e) Comparison of model-based ($k_{t}$; solid) and experimental ($k_{exp}$; dashed) permeability dynamics for \( \Delta \)fliC (blue) and WT (red) strains. The only fitting parameter is $k_{\mathrm{bf}}$.}
    \label{Fig5}
\end{figure*}

\noindent
The overall porous structure is conceptualized as a series of $m$ virtual porous elements, as shown in Fig.~\ref{Fig5}~$a$. Each of these porous elements is characterized by a given length, $l_i$, along the mean flow direction and permeability $k_i$ ($i = 1, \dots, m$). The total number of porous elements required to span the entire domain is $m = L / l$, where $L=44.8$~mm is the total system length and $l$ the average porous element size. In the following, we directly relate permeability $k_i$ and length $l_i$ of each element to the structural properties of the medium. The porous element extent is the pore size $l_i = D_{mic_{i}} / \tau$ measured via Maximum Inscribed Circle, as discussed above (Fig.~\ref{Fig5}~a and Supp. Info.) projected along the mean flow path ($\tau = 1.4$ being the medium tortuosity~\cite{jiao2024intrinsic}). Each virtual element corresponds to an individual pore. Hydraulically, a fully saturated pore behaves like a pipe whose permeability is governed by the Hagen-Poiseuille formulation. Biofilm growth within a pore alters its hydraulic behavior in two main ways: ($i$) reducing the area available for fluid flow; and ($ii$) being an additional porous medium, coating the pore walls within. Under pressure imposed conditions and injecting sterile nutrient solution, we do not observe the formation of streamers~\cite{rusconi2010laminar, drescher2013biofilm, scheidweiler2019unraveling}. In this framework, the number of porous elements is $m = L \tau / \overline{D_{MIC}} = 1334$. \\

\noindent
The flow in each pore \(i\) is represented as fluid moving through two parallel flow systems: a central pipe (representing the biofilm-free region) and an outer (annular) biofilm layer. The central biofilm-free pipe has diameter \(d_{1_i}\) and permeability given by Hagen-Poiseuille law \(k_{\mathrm{d}_{1_i}} = d_{1_i}^2/32\), while the annular biofilm of thickness \(d_{2_i}/2\) has permeability \(k_{\mathrm{bf}}\), as represented schematically in Fig.~\ref{Fig5}~c. In each pore, we quantify the biofilm layer of thickness as follows. We measure i) the fraction of the pore area occupied by biofilm $A_{{\mathrm{mic}}_i}$ relative to the total pore area and ii) the bacterial fluorescent signal \(I_{\mathrm{mic}}\), to define the biomass density as $\rho_{{pi}_i} = \frac{\sum I_{{\mathrm{mic}}_i}}{I_{{\mathrm{max}}} A_{{\mathrm{mic}}_i}}$. The temporal dynamics of the latter embeds dynamic change in pore space due to biomass accumulation or removal (see also methods). Therefore, the biomass thickness is quantified as \(d_{2i}(t) = \rho_{\mathrm{pi}_i}(t) D_{i{\text{MIC}}}/I_{\mathrm{max}}\). Thus, the equivalent permeability of the $i-th$ pore \(k_{ei}(t)\) is evaluated as the combined contribution of these these two parallel flow paths:
\begin{equation}
    k_{ei}(t) = \frac{d_{1i}(t) k_{\mathrm{d}_{1i}}(t)  + d_{2i}(t) k_{\mathrm{bf}}}{d_{1i} (t) + d_{2i}(t) }.
\end{equation}

\noindent
We further analyze all pores in the system (approximately 105,000 across the entire structure) and quantify the statistical correlation between pore size and biomass density. Our results suggest that the two variable are statistically independent (see Supplementary Fig.~S9). In other words, we find large and small pores with either high or low biomass density. This observation can be justified by the fact that pore system considered is heterogeneous and spatially extended ($L \gg \overline{D_{mic}}$). Therefore, large and small pores can be found in high velocity as well as in fluid stagnation zones, thus, bacteria in all pores can be exposed to high or low resources concentrations, leading to high or low biomass accumulation. To quantify the impact of the biomass accumulation on pore space and overall permeability, we assume that the macroscopic porous domain consists of $m = L \tau / \overline{D_{MIC}} = 44.8 \cdot 1.4 / 0.047 = 1334$ independent pores. We, then, randomly pick $m$ pore sizes $D_i$ sampled from the measured distribution of pore diameters (from Maximum Inscribed Circle analysis, shown in Fig.\ref{Fig1}~$f$) and, independently, a biofilm density $\rho_{i,\text{bf}}$ sampled from the experimental biomass density distribution measured at time $t$ (see Supplementary Fig.~S5~$a,b$). From these values we estimate the biomass thickness $d_{2i}$ and the biomass-free pipe diameter $d_{1i}$ and permeability. Finally, the overall medium permeability $k_t(t)$ is evaluated as the harmonic mean of the corresponding permeability of each individual element ($1, ..., m$), as~\cite{jiao2024intrinsic}:
\begin{equation}\label{K_model}
    k_t(t) = \frac{L\tau^2}{\sum \frac{D_{i,MIC}}{k_{ei}(t)}} = \frac{L\tau^2}{\sum \frac{D_{i,MIC}(t) (d_{1i} (t) + d_{2i}(t))}{d_{1i}(t) k_{\mathrm{d}_{1i}}(t) + d_{2i}(t) k_{\mathrm{bf}}}}.
\end{equation}
Note that biofilm permeability, $k_{bf}$, is the only parameter estimated in our model, all other quantities being directly measured. Here, $k_{bf}$ is interpreted as an effective permeability of a biofilm-occupied pore that is slightly patchy, or characterized by channels / gaps. Although variations in EPS composition, compaction, and viscoelastic properties may, in principle, alter biofilm permeability over time, the fact that a single $k_{bf}$ value consistently captures both WT and $\Delta$fliC experiments indicates that biomass spatial organization is the dominant control on permeability under the present experimental conditions. This assumption is further supported by the WT case, where the system permeability remains stable across all three replicas beyond the characteristic time $t_c = 7.5$~h. As illustrated in Fig.~\ref{Fig5}~$d,e$, the resulting theoretical permeability model $k_t$ is fully consistent with the measured permeability $k_{\text{exp}}$ for the motile and non-motile bacterial populations (see Supplementary Information for all replicas). The biofilm permeability is estimated as the best parameter value to fit the data in Fig.~\ref{Fig5}~$d,e$ and results to be $k_{bf} = 2.5$~darcy in both scenarios. This value falls within the range of values previously documented in the literature~\cite{kurz2023morphogenesis, gaol2021investigation}. \\

\subsection*{Growth-limiting mechanisms control permeability predictions}

\noindent
To complement the theoretical framework described above, we consider the prediction of the developed model to biofilm accumulation patterns that could emerge under two contrasting growth regimes: nutrient-limited and space-limited conditions. These growth regimes are governed by distinct biological and environmental constraints and reasonably lead to fundamentally different statistical distributions of biofilm density inside each pore, defined as $\rho = \mathrm{BM} / A$ (where $\mathrm{BM}$ denotes biomass contained in a single pore of area $A$). To investigate the way growth constrains and influences the spatial distribution of biomass in these two scenarios, we implement a logistic growth model~\cite{madigan2006brock} in $m = L/l$ pores that fluid has to cross. In each individual pore, the logistic model is controlled by two parameters: the growth rate $r$ and the carrying capacity $K$~\cite{madigan2006brock}. Normalizing time by the characteristic growth time needed for the bacterial population to reach half the carrying capacity ($BM(t=t_{1/2} = K/2$), the growth rate $r$ does not play any role in the overall dynamics. We define a homogeneous (single value) carrying capacity $K$ for a nutrient-limited scenario: in other words, regardless the pore size bacteria can grow until nutrients are depleted. In contrast, for a space-limited scenario, we define a pore-dependent carrying capacity proportional to the individual pore area $A$, $K \propto A$: in other words, bacteria can grow in each pore until there is no more space to host new cells. To understand the coupling between growth conditions and medium structure, we consider these two scenarios in heterogeneous media (pores of distributed size) and in a homogeneous porous medium, where all pores have the same size. \\

\begin{figure*}[htb!]
\centering
\includegraphics[width=1\linewidth]{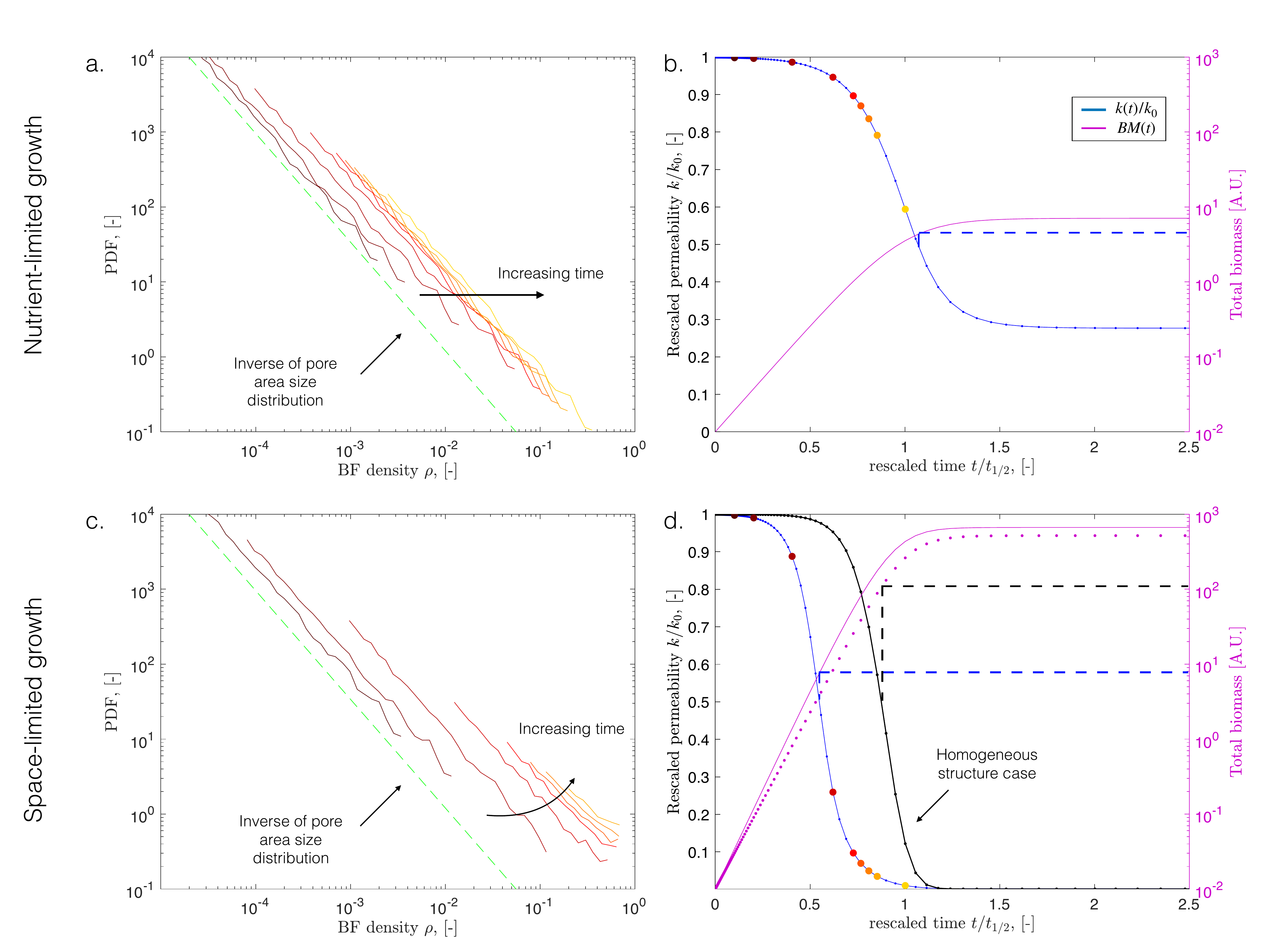}
\caption{Prediction of permeability decrease for nutrient-limited and space-limited growth conditions in heterogeneous pores. (a, c) Probability density functions (PDFs) of biofilm density ($\rho_{pi}$) at different times for nutrient-limited (a) and space-limited (c) regime. Dashed line represents the distribution of pore sizes $D_{MIC}$. In the first case we observe a persistence of skewed distribution (a) while in the second case it emerges a near-Gaussian symmetric pattern (c). The dashed line represents the PDF of the pore size. (b, d) Dynamics of the overall biomass ($\bar{\rho}_{pi}(t)$) and permeability prediction $k_t(t)$, eq.~\eqref{K_model} under nutrient-limited (b) and space-limited (d) regimes. Time is normalized by $t_{1/2}$, the time at which the biomass reaches half of its carrying capacity.}
\label{Fig6}
\end{figure*}

\noindent
\textbf{Early times}. Under both nutrient- and space-limited regimes, the early-time distribution of \( \rho \) predicted by integrating the biomass across all pores is primarily controlled by the pore size. Starting from a uniform initial biomass $BM_0$ (every pore has the same biomass) and proceeding at same growth rates across pores, $BM$ increases to the same extent in every pore (without filling entirely any pore), yielding $\rho \propto 1/A$. This behavior, shown in Fig.~\ref{Fig6}~$a,c$ by the solid lines of darker color (earlier times) is consistent with our empirical observations at early stages of growth (see Fig.~\ref{Fig3} and Supplementary Fig.S5). \\

\noindent
\textbf{Later times}. Under nutrient-limited conditions, typical of low-nutrient environments~\cite{hornung2018quantitative, kim2000biomass}, biomass accumulation halts uniformly across the porous network once local resources are depleted. Given that nutrient accessibility is independent of pore size (as slow growth ensure homogeneous and rapid nutrient refreshing by flow and diffusion), biomass in each pore attains the carrying capacity $K$, such that $BM \approx K$. Consequently, biofilm density keeps scaling inversely with $\rho \propto 1/A$, thus mirroring pore area distribution. Since pore sizes in natural porous structures are often well approximated by multi-scale (e.g. gamma)  distributions~\cite{miele2019stochastic, jiao2024intrinsic}, this mechanism inherently leads to a broad distribution of biofilm density values as shown in Fig.~\ref{Fig6}~$a$ (for experiments see Supplementary Fig.~S5~a-c-e). 
By contrast, under space-limited conditions, typical of fast-growing strains or nutrient-rich environments, nutrients remain accessible throughout the growth process, and the physical confinement imposed by pore geometry becomes the dominant constraining factor. In these conditions, once a pore is filled with biomass its growth stops, thus, $BM$ scales with the pore size, (i.e., $BM \propto A$, in other words more space, more biomass), leading towards a uniform and narrow distribution for biofilm density values $\rho = BM / A \propto \textrm{const.}$. This is shown in Fig.6~$c$, where the biomass density distribution is initially broad and scales as $1/A$ while becoming narrower at later times (for experiments see Supplementary Fig.~S5~b-d-f).\\

\noindent
Fig.~\ref{Fig6}~$b,d$ shows as purple solid line (semi-logarithmic plot) the overall growth obtained by summing the biomass in all pores, versus time rescaled by the characteristic growth time $t_{1/2}$. At early times the biomass accumulation increases exponentially fast for all scenarios. As soon as the rescaled time approaches $t/t_{1/2} = 1$, the $BM$ growth slows down and stabilizes over a constant value (macroscopic carrying capacity). For the nutrient-limited case this value is much smaller (about 100 times) than for the the space-limited one: this is expected since, for the first case, we imposed a single, same for each pore, and small carrying capacity while for the second case the carrying capacity is pore-dependent (i.e. for large pores it is larger). \\

\noindent
Building on these elements, we integrate the obtained biomass density distributions into our (biomass-controlled) permeability model of eq.~\eqref{K_model} to estimate the dynamics of macroscopic hydraulic properties. The resulting simulation outcomes are shown in Fig.~\ref{Fig6}~$b,d$. Under nutrient-limited conditions, once the total biomass reaches half of the carrying capacity (at time $t=t_{1/2}$), the system permeability decreases to about 50\% of its initial value, as highlighted by the blue dashed line (Fig.~\ref{Fig6}~$b$). It, then, reduces further down to 30\% of its initial value as biomass continues accumulating until carrying capacity is attained. Otherwise, under space-limited conditions, for the permeability to drop to about 50\% of its initial value, it is required a much shorter time ($t/t_{1/2} = 0.5$) and a similar overall biomass (Fig.~\ref{Fig6}~$d$). Then, the permeability continues to decrease as biomass further increases, until all pores become fully clogged, reducing to almost zero $t / t_{1/2} > 1$. At that point, the effective (system-scale) medium permeability is entirely controlled by the biomass occupying the entire pore space ($k_{bf}$). \\

\noindent
As a control, we also simulated logistic biomass growth in a homogeneous porous system for which the carrying capacity is set proportional to the pore size. All pores are identical to the average pore size of the heterogeneous case considered: thus, there is no difference between nutrient- and space-limited scenario. The result is shown in Fig.~\ref{Fig6}~$d$. The overall biomass growth (purple dots) is similar to the space-limited scenario in the heterogeneous medium (purple solid line), however, the permeability decay is remarkably different (black dotted line). For the permeability to decrease down to about 50\% its own initial value, it is required about 20 times the biomass that clogged the heterogeneous medium, requiring also more time. Thus, we conclude that the growth regime, shaping biomass accumulation and spatial organization, together with the structure of the host porous system control the macroscopic medium permeability. \\

\noindent
Our study elucidates how biomass growth dynamically alters intrinsic permeability in heterogeneous porous media. We document that biomass spatial organization (rather than total biomass alone) controls permeability evolution, and that this organization is strongly influenced by bacterial motility. Motile strains redistribute in response to resource gradients, leading to reduced downstream accumulation where nutrients are depleted. In contrast, non-motile strains remain localized and continue to grow under resource-limited conditions, progressively filling pores and promoting stronger clogging. We develop a model that quantitatively captures the observed permeability dynamics by linking pore-scale biomass distributions to the macroscopic hydraulic response. The model requires spatio-temporal biomass accumulation and, therefore, reflects the integrated outcome of local biomass growth, redistribution, and loss processes. Within this framework, we show that spatial organization of biomass within the pore space is a key control on permeability under the present experimental conditions. \\

\noindent
Our findings provide new insights into biomass-mediated permeability alteration in porous systems with implications for applications such as water treatment, filtration, enhanced oil recovery, and soil bioremediation. While the experiments are performed in a controlled quasi-two-dimensional porous system, the conceptual framework is not limited to two dimensions. Extension to three-dimensional media would require appropriate pore-scale descriptors (e.g., maximum inscribed balls) and corresponding measurements of biomass occupancy. Future work should incorporate explicit nutrient transport and growth dynamics, as well as explore flow-controlled conditions where shear-induced detachment further modulates permeability dynamics.\\

\noindent
\textbf{Supporting Information}\\
Additional methodological and modeling details are described in the Supporting Information file; including i) pressure control system, ii) UV-C irradiation setup, iii) biomass growth quantification, iv) flow experiment procedures, and v) permeability model validation; supplementary figures show i) temporal permeability and biomass evolution for al replicas, ii) longitudinal and time-lapse biomass distributions, iii) biofilm-density probability distributions, iv) pressure and flow-rate controls, and v) \(D_{\mathrm{MIC}}\)--\(\rho_{pi}\) relationships for \(\Delta \mathrm{fliC}\) and WT strains. \\

\noindent
\textbf{Data Availability}\\
The collected experimental data, simulation results and codes, and the plots data used and discussed in this study are available upon request. \\

\noindent
\textbf{Competing Interests}\\
The authors declare no competing interests. \\

\noindent
\textbf{Acknowledgments}\\
P.d.A. acknowledges the support of FET-Open project NARCISO (ID: 828890) and the Swiss National Science Foundation (grants ID~200021$\_$172587 and 200021$\_$219863). W.J. acknowledges the funding of China Scholarship Council for financial support through the fellowship (grant ID: CSC202008210309). AG acknowledges support from the European Union Next-Generation EU (National Recovery and Resilience Plan - NRRP, Mission 4, Component 2, Investment 1.3 - D.D. 1243 2/8/2022, PE0000005) in the context of the RETURN Extended Partnership.\\

\noindent
\textbf{Author Contributions}\\
W.J. and P.d.A. designed the research with input form N.W. and D.S., W.J. performed experiments with input form D.S.; W.J. and N.W. developed the feedback loop for pressure control system, W.J. and P.d.A. analyzed the data, W.J., A.G. and P.d.A. derived the theoretical model and all authors wrote the manuscript. \\

% Bibliography
%\subsection**{References}
%\bibliography{library}


\end{document}